\documentclass[journal]{IEEEtran}
\usepackage{epsfig,amsfonts,amsbsy,bm,mathrsfs}
\usepackage[nolist]{acronym}
\usepackage{mathrsfs} 
\usepackage{graphicx,cite,amssymb,amsmath,bm}
\usepackage{mathtools}
\usepackage{dsfont}
\mathtoolsset{showonlyrefs}
\usepackage{amsmath} 
\usepackage[utf8]{inputenc}
\usepackage{subcaption}
\usepackage{tikz}
\usetikzlibrary{fadings}
\usetikzlibrary{shadows.blur}
\usetikzlibrary{shapes,arrows}
\usetikzlibrary{calc,shapes.misc}
\usetikzlibrary{decorations.pathreplacing}
\usepackage{ifthen}
\usetikzlibrary{positioning}
\usepackage{pgfplots,relsize}
\usetikzlibrary{plotmarks}
\usepackage{ctable}
\usepackage{pgfplots}
\usepackage[english]{babel}
\usepackage{xcolor}
\usepackage{color}
\usepackage[utf8]{inputenc}
\usepackage[T1]{fontenc}
\usetikzlibrary{decorations.pathreplacing,shapes.misc}
\usetikzlibrary{patterns}
\usepackage{colortbl}
\usepackage{balance}
\usepackage{perpage}
\MakePerPage{footnote}

\begin{document}

\renewcommand{\mathbf}{\bm}
\newcommand{\argmax}[1]{\underset{#1}{\mathrm{arg \, max}}\,}
\newcommand{\argmin}[1]{\underset{#1}{\mathrm{arg \, min}}\,}
\newcommand{\code}{\mathcal{C}}
\newcommand{\mm}{\mathrm{q}}
\newcommand{\mi}{{\mathrm{I}}}
\newcommand{\inalpha}{\mathcal{X}}
\newcommand{\outalpha}{\mathcal{Y}}
\newcommand{\inx}{\mathsf{X}}
\newcommand{\outy}{\mathsf{Y}}
\newcommand{\modcw}{\mathbf{x}}
\newcommand{\obs}{\mathbf{y}}
\newcommand{\rate}{R}
\newcommand{\coderate}{r}
\newcommand{\modblocklen}{N}
\newcommand{\codblocklen}{n}
\newcommand{\cw}{\mathbf{c}}
\newcommand{\M}{M}
\newcommand{\Bmat}{\mathbf{B}}

\newcommand{\hb}{h_\mathrm{b}}
\newcommand{\es}{\delta}
\newcommand{\eb}{\epsilon}
\newcommand{\x}{\boldsymbol{x}}

\newcommand{\pyx}{p_{Y|X}}
\newcommand{\phxx}{P_{\hat{X}|X}}
\newcommand{\de}{\mathrm{d}}
\newcommand{\avg}{\mathbb{E}}
\newcommand{\cfield}{\mathbb{C}}
\newcommand{\hd}{d_{\mathrm{H}}}

\newcommand{\expect}{\mathbb{E}}

\newcommand{\SNR}{\mathsf{SNR}}

\newcommand{\HDDSW}{\mathsf{HDD-SW}}
\newcommand{\HDDBW}{\mathsf{HDD-BW}}

\newcommand{\otoprule}{\midrule[\heavyrulewidth]}

\definecolor{mygreen}{rgb}{0.0, 0.45, 0.0}

\newtheorem{mydef}{Definition}
\newtheorem{prop}{Proposition}
\newtheorem{theorem}{Theorem}
\newtheorem{lemma}{Lemma}
\newtheorem{example}{Example}
\newtheorem{corollary}{Corollary}

\newcommand{\rf}{\textcolor{blue}}
\newcommand{\rs}{\textcolor{red}}
\newcommand{\rt}{\textcolor{mygreen}}
\newcommand{\AS}{\textcolor{green}}
\begin{acronym}
	\acro{AWGN}{additive white Gaussian noise}
	\acro{BSC}{binary symmetric channel}
	\acro{CM}{coded modulation}
	\acro{CSI}{channel state information}
	\acro{DMC}{discrete memoryless channel}
	\acro{FEC}{forward error correction}
	\acro{HD}{hard decision}
	\acro{i.i.d.}{independent and identically distributed}
	\acro{MAP}{maximum a posteriori}
	\acro{ML}{maximum likelihood}
	\acro{r.v.}{random variable}
	\acro{RCB}{random coding bound}
	\acro{SD}{soft decision}
	\acro{SSD}{soft-detection/decoding}
	\acro{HSD}{hard-detection/soft-decoding}
	\acro{HHD}{hard-detection/decoding}
	\acro{BHSD}{bit-wise hard-detection/soft-decoding}
	\acro{BHHD}{bit-wise hard-detection/decoding}
	\acro{MHHD}{multilevel hard-detection/decoding}
	\acro{GMI}{generalized mutual information}
	\acro{QAM}{quadrature amplitude modulation}
	\acro{SNR}{signal-to-noise ratio}
\end{acronym}

%%%%%%%%%%%%%%%%%%%%%%%%%%%%%%%%%%%%%%%%%%%%%%%%%%%%%%%%%%%%%%%%%%%
\title{Achievable Information Rates for Coded Modulation with Hard Decision Decoding for Coherent Fiber-Optic Systems}
\author{
	Alireza Sheikh \IEEEmembership{Student Member, IEEE}, Alexandre Graell i Amat, \IEEEmembership{Senior Member, IEEE}, \\and Gianluigi Liva, \IEEEmembership{Senior Member, IEEE}
	\thanks{Part of this paper will be presented at the European Conference on Optical Communications (ECOC), Gothenburg, Sweden, 2017 \cite{Sheikh17}.}
	\thanks{This work was funded by the Knut and Alice Wallenberg Foundation and the Swedish Research Council under grant 2016-04253.}
	\thanks{A. Sheikh and A. Graell i Amat are with the Department of Electrical Engineering, Chalmers University of Technology, SE-41296 Gothenburg, Sweden (email: {asheikh, alexandre.graell}@chalmers.se).}
	\thanks{G. Liva is with the Institute of Communications and
		Navigation of the German Aerospace Center (DLR), M\"unchner Strasse 20, 82234 We{\ss}ling, Germany (email: gianluigi.liva@dlr.de).}}

\maketitle
\thispagestyle{empty}

\begin{abstract}
	
	We analyze the achievable information rates (AIRs) for coded modulation schemes with QAM constellations with both bit-wise and symbol-wise decoders, corresponding to the case where a binary code is used in combination with a higher-order modulation using the bit-interleaved coded modulation (BICM) paradigm and to the case where a nonbinary code over a field matched to the constellation size is used, respectively. In particular, we consider hard decision decoding, which is the preferable option for fiber-optic communication systems where decoding complexity is a concern. Recently, Liga \emph{et al.} analyzed the AIRs for bit-wise and symbol-wise decoders considering what the authors called \emph{hard decision decoder} which, however, exploits \emph{soft information} of the transition probabilities of discrete-input discrete-output channel resulting from the hard detection. As such, the complexity of the decoder is essentially the same as the complexity of a soft decision decoder. In this paper, we analyze instead the AIRs for the standard hard decision decoder, commonly used in practice, where the decoding is based on the Hamming distance metric. We show that if standard hard decision decoding is used, bit-wise decoders yield significantly higher AIRs than symbol-wise decoders. As a result, contrary to the conclusion by Liga \emph{et al.}, binary decoders together with the BICM paradigm are preferable for spectrally-efficient fiber-optic systems. We also design binary and nonbinary staircase codes and show that, in agreement with the AIRs, binary codes yield better performance.
\end{abstract}


\section{Introduction}\label{sec:intro}

\IEEEPARstart{T}{he} ever increasing demand in network capacity has made the adoption of forward error correction (FEC) a must in optical communications. To achieve high spectral efficiencies, a common approach is to use a powerful binary code followed by a nonbinary constellation, exploiting the bit interleaved coded modulation (BICM) paradigm. An alternative is to use a nonbinary code, matched to the constellation size.

The analysis and design of binary low-density parity-check (LDPC) codes and spatially-coupled LDPC (SC-LDPC) codes, in combination with higher order modulation and soft decision decoding (SDD), has recently attracted a great deal of attention in the optical communications community \cite{Djordjevic2004,Djordjevic_GLDPC,Hager15,Schmalen15}. FEC with SDD yields very large coding gains, but poses implementation challenges in terms of complexity, delay, and power consumption at very high data rates. This motivates the use of FEC schemes based on algebraic codes, which are decoded using the less complex hard decision decoding (HDD). Binary staircase codes \cite{Smith12,Hager15b}, braided codes \cite{PfisterGC13}, and generalized product codes \cite{Hag16b} have been shown to perform close to the theoretical limits while achieving the very low error rates needed in optical communications. Nonbinary staircase codes with HDD were considered in \cite{She17}.

Useful parameters to determine the ultimate performance limits of these so-called coded modulation (CM) schemes and compare them, are achievable information rates (AIRs), which provide a lower bound on the mutual information (MI) of the system, i.e., the maximum rate at which reliable communication is possible. The fiber-optic channel is characterized by memory. In \cite{Djordjevic2005}, AIRs for high-speed optical transmission with on-off keying and symbol-wise SDD were derived by modeling the joint effect of nonlinearity and dispersion as a finite-state machine \cite{arnold2006}. However, the complexity of the computation increases for higher order modulation due to an increased number of states. In \cite{Ess10}, Essiambre \emph{et al.} obtained capacity lower bounds for a variety of scenarios using a channel model based on extensive lookup tables. An alternative approach to compute AIRs is to resort to \emph{mismatch decoding} \cite{Kaplan1993} and to neglect the memory of the channel, leading to a lower bound on the MI \cite{Sec13,Fehenberger_15,Lig16}. In \cite{Fehenberger_15}, AIRs for SDD and bit-wise HDD assuming mismatch decoding were computed for long-haul fiber-optic systems where the effect of transmit power was studied for different compensating methods. In \cite{Lig16}, AIRs for several CM schemes were derived. In particular, the authors analyzed AIRs for CM schemes with bit-wise and symbol-wise decoding (suitable for binary and nonbinary codes, respectively) considering both SDD and what the authors referred to as HDD. A significant outcome of the analysis in \cite{Lig16} is that nonbinary codes with HDD can achieve information rates comparable to that of SDD, and quoting \cite{Lig16} ``hard decision binary decoders are shown to be unsuitable for spectrally-efficient, long-haul systems''.

In this paper, we further elaborate on the AIRs of CM for HDD using binary and nonbinary codes. Our analysis extends the results in \cite{Lig16} and provides new insight, clarifying some concepts and shedding some light into the conclusions in \cite{Lig16}. In particular, we remark that the AIRs that the authors computed in \cite{Lig16} for HDD corresponds to the AIRs of a CM scheme where the detector takes a hard decision on the channel output, i.e., it performs \emph{hard detection}, but the decoder exploits the transition probabilities of the resulting discrete-input discrete-output channel \cite[eq.~(23)]{Lig16}. In other words, the decoder uses a metric that carries \emph{soft information} about the reliability of the demodulator hard decisions. %Whether a decoder employing this metric can be referred to as a hard decision (HD) decoder is an open question, which depends on the definition of HDD. 
According to \cite{LinCos04,lapidoth17,barg1997complexity,guruswami2004list,vardy1997algorithmic}, a hard decision decoder decides solely based on the Hamming distance, hence the decoder of \cite{Lig16} does not fall within this category. Therefore, to avoid ambiguities we will refer in the following to the decoder in \cite{Lig16} as hard detector/channel-aware (HdCha) decoder, and to the (standard) hard decision (HD) decoder that uses the Hamming distance metric \cite{LinCos04,lapidoth17,barg1997complexity,guruswami2004list,vardy1997algorithmic} simply as the HD decoder. Indeed, it is important to remark that the standard HD decoder allows for the use of low-complexity algebraic decoding algorithms and therefore is the one used in practice, e.g., to decode Reed-Solomon (RS) codes, Bose-Chaudhuri-Hocquenghem (BCH) codes, and staircase codes. On the other hand, decoders implementing the metric in \cite[eq.~(23)]{Lig16} may be assimilated to soft decoders with coarse quantization and it is unclear whether a low-complexity implementation is possible.

For (standard) HDD, a relevant AIR is not the one considered in \cite{Lig16}, but the one that assumes the use of the Hamming decoding metric. In this paper we therefore derive an AIR of a CM scheme with HDD using the Hamming metric for both bit-wise and symbol-wise decoders. We then compare AIRs for SDD, HDD, and the HdCha decoding (HdChaD) of \cite{Lig16}, assuming both bit-wise and symbol-wise demapper. As expected, AIRs for HDD are significantly lower than the ones for the HdChaD scheme considered in \cite{Lig16}. Furthermore, we show that the AIR of HDD with bit-wise metric is higher than that of the CM scheme with symbol-wise metric. Indeed, the AIR of the CM scheme with symbol-wise HDD is much worse than that of the scheme with SDD. Therefore, interestingly, contrary to the conclusion in \cite{Lig16}, if a standard HD decoder is used, binary codes and BICM are to be preferred to nonbinary codes for spectrally-efficient optical communication systems.

We also consider binary and nonbinary staircase codes with HDD and compare their asymptotic performance (i.e., for very large block lengths), derived using density evolution (DE), and their finite block length performance to the obtained AIRs for the \ac{AWGN} channel. The results are in agreement with the AIRs and confirm that, when HDD is used, binary codes are preferable to nonbinary codes. Finally, we corroborate this outcome by analyzing AIRs of the fiber-optic channel. In particular, we compute AIRs of the polarization-multiplexed (PM) single channel transmission system using the split-step Fourier method (SSFM) and of the wavelength division multiplexing (WDM) system using the Gaussian noise (GN) model. We show that for HDD, binary codes/bit-wise decoders yield a significant optical reach enhancement compared to nonbinary codes/symbol-wise decoders.
This paper extends our previous work \cite{Sheikh17}.\\
Notation: The following notation is used throughout the paper. We denote by $\mathbb{C}$ the set of complex numbers. We use boldface letters to denote vectors, e.g., $\x$, and capital letters to denote random variables, e.g., $X$. The cardinality of a set $\mathcal{A}$ is denoted by $|\mathcal{A}|$.

\section{Preliminaries}\label{sec:prelim}

We consider the memoryless AWGN channel, which is an accurate model for long-haul coherent fiber-optic communications when the fiber-optic channel is dominated by amplified spontaneous emission (ASE) noise \cite{Pog12}. The transition probability density is 
\begin{align}
\pyx(y|x)=\frac{1}{2\pi\sigma^2}\exp\left[-\frac{1}{2\sigma^2}\lvert y-x\rvert^2\right]
\end{align}
with $x\in \inalpha$, where $\inalpha=\left\{\inx_1, \inx_2,\ldots,\inx_\M \right\}\subset \cfield$ is the input alphabet (i.e., the constellation imposed by the  modulation), and where $y\in \cfield$ is the channel output and $\sigma^2$ is the noise variance. We consider the case where the input alphabet is restricted to a \ac{QAM}. We define the \ac{SNR} as $\SNR=E_{\mathsf s}/N_{\mathsf 0}=1/2\sigma^2$, $E_{\mathsf s}$ being the average energy per symbol and $N_{\mathsf 0}$ the single-sided noise power spectral density. We assume a coded transmission where $K$ information symbols (i.e., $k=K\log_2 M$ information bits) are encoded into $N$ modulated symbols by an $(N,K)$ code $\code$ of rate $R=K/N$. Thus, spectral efficiency of the systems is $R \log_2 M$ [bpcu]. 

We consider several detection/decoding strategies for both both symbol-wise and bit-wise decoders.

\subsection{Coded Modulation with Symbol-Wise Soft Decision Decoding}\label{sec:prelim_CMSD}

For symbol-wise SDD (SDD-SW), the detector provides the FEC decoder with the likelihoods $\pyx(y|x)$ for all $x \in \inalpha$. The FEC decoder then uses the \ac{ML} decoding rule,  given by
\begin{equation}\label{SSD_detect}
\hat{\modcw}_{\mathsf{SSD-SW}}=\argmax{\modcw \in \code} \prod_{i=1}^{\modblocklen}\pyx(y_i|x_i),
\end{equation}
where $\modcw$ is the codeword of modulated symbols and $\hat{\modcw}$ is the decoded codeword.
%\rf{where $\code$ is the codebook.}

\subsection{Coded Modulation with Symbol-Wise Hard-Detection Channel-Aware Decoding}\label{sec:prelim_CMHDSD}

We now consider the symbol wise HdChaD (HdChaD-SW) scheme in \cite{Lig16}. With HdChaD-SW, the detector takes a symbol-wise hard decision, 
\begin{equation}\label{H_detect}
\hat{x}_i=\argmax{x \in \inalpha} \pyx(y_i|x).
\end{equation}
The decision is then passed to the FEC decoder. By doing so, the channel is turned into an $M$-ary input, $M$-ary output \ac{DMC} with transition probabilities
\begin{align}
\phxx(\hat{x}|x)=\Pr\left\{\hat{X}=\hat{x}|X=x\right\}.\label{eq:DMC}
\end{align}
Note that the output alphabet of the \ac{DMC} is $\inalpha$.
The optimum decoding rule is given by
\begin{align}
\label{eq:HdChaD}
\hat{\modcw}_{\mathsf{HdChaD-SW}}=\argmax{\modcw \in \code} \prod_{i=1}^{\modblocklen}\phxx(\hat{x}_i|x_i).
\end{align}
Observe that, although the detector takes a hard decision, the FEC decoder is not a standard HD decoder, in the sense that the decoding metric still embeds soft information. In fact, the use of the channel transition probabilities $\phxx(\hat{x}|x)$ of the DMC implies that the decoder is aware of the geometry of the constellation (and eventually of the channel SNR), and exploits this information to obtain likelihoods for the different channel input-output pairs according to \eqref{eq:DMC}. In this sense, the detector can be seen as a soft detector with a (coarse) quantization. 

The discussion above highlights that hard detection is not equivalent to HDD. In particular, hard detection does not imply HDD: Even if hard detection is performed, the channel transition probabilities of the resulting DMC are soft information that can be exploited by a SDD. As explained in \cite{Schmalen2017}, the hard-detector channel-aware decoder that implements the decoding rule \eqref{eq:HdChaD} can be implemented by computing log-likelihood ratios based on the channel transition probabilities of the resulting DMC and feeding them to a conventional soft decision decoder. However, the decoding complexity is then the same as that of SDD.

\subsection{Coded Modulation with Symbol-Wise Hard Decision Decoding}\label{sec:prelim_CMHDHD}
We now consider a CM scheme with the conventional symbol-wise HDD (HDD-SW). Also in this case the detector takes a symbol-wise hard decision,
\[
\hat{x}_i=\argmax{x \in \inalpha} \pyx(y_i|x).
\]
The decision is passed to the FEC decoder, which uses the Hamming distance as decoding metric. With optimum Hamming-metric decoding we have
\[
\hat{\modcw}_{\HDDSW}=\argmin{\modcw \in \code} \hd\left(\hat{\bm{x}}, \modcw \right)
\]
with the Hamming distance defined as 
\[
\hd\left(\hat{\x}, \modcw \right)\triangleq\lvert \left\{i|\hat{x}_i\neq x_i \right\}\rvert.
\]
Observe that the Hamming decoding metric is the one that is conventionally assumed for HDD (enabling low-complexity decoding, see, e.g., the Berlekamp-Massey algorithm for RS and BCH codes) \cite{LinCos04,lapidoth17,barg1997complexity,guruswami2004list,vardy1997algorithmic}. We also remark that the bounded distance decoding (BDD), which is the low-complexity decoding algorithm used to decode staircase codes and GPCs, utilizes the Hamming metric, i.e., the received word is decoded to a codeword if the Hamming distance between the received word and a codeword is less or equal to the error correcting capability of the code (see \cite[Sec. II]{Henry_approachingcapacity_12}). 

\section{Achievable Information Rates for Symbol-Wise Decoding and Uniform Input Distribution}\label{sec:bounds}
We consider next the achievable information rates for the three schemes introduced in the previous section, for the case of uniform input distribution, i.e., $p_X(x)=1/M$ for all $x\in{\inalpha}$.

\medskip

\subsection{Achievable Information Rate for Symbol-Wise Soft Decision Decoding}\label{sec:ISDD}

Under SDD-SW, the achievable information rate is given 
\begin{align}
I_{\mathsf{SDD-SW}}&=I(X;Y)\\
&=\expect\left[ \log_2 \left( \frac{\pyx(Y|X)}{\frac{1}{M}\sum_{x\in\inalpha}\pyx(Y|x)}\right)  \right]
\end{align}
where the expectation is over $X,Y$.

\subsection{Achievable Information Rate for Symbol-Wise Hard Detection Channel-Aware Decoding}

For HdChaD-SW, the achievable information rate is given by \cite[eq. 23]{Lig16}
\begin{align}
I_{\mathsf{HdChaD-SW}}&=I(X;\hat{X})\\
&=\expect\left[ \log_2 \left( \frac{\phxx(\hat{X}|X)}{\frac{1}{M}\sum_{x\in\inalpha}\phxx(\hat{X}|x)}\right)  \right]
\end{align}
where the expectation is over $X,\hat{X}$.

\subsection{Achievable Information Rate for Symbol-wise Hard Decision Decoding}
\label{sec:AIRsHDDsw}

Under HDD-SW, a possible strategy to compute an achievable information rate is to resorting to the \emph{mismatched decoding} framework. It is easy to observe that employing the Hamming decoding metric is equivalent to maximizing the mismatched metric
\begin{equation}
\mm(\hat{\bm{x}},\modcw)=\prod_{i=1}^N \mm(\hat{x}_i,x_i)
\end{equation}
with
\begin{align}
\mm(\hat{x},x)=\left\{
\begin{array}{ll}
1-\epsilon & \qquad \text{if }\hat{x}=x\\
\epsilon/(M-1) & \qquad \text{otherwise}
\end{array}\right. \label{eq:MM_symbol}
\end{align}
where $\epsilon$ is an arbitrary value in $(0,(M-1)/M)$.
In fact,
\begin{align}
\hat{\modcw}_{\HDDSW}&\,=\argmax{\modcw \in \code} \prod_{i=1}^{\modblocklen}\mm(\hat{x}_i,x_i)\\ &\,=\argmax{\modcw \in \code} \left[\frac{\epsilon}{(1-\epsilon)(M-1)}\right]^{\hd\left(\hat{\modcw}, \modcw \right)}\\[2mm]
&\stackrel{(a)}{=}\argmin{\modcw \in \code} \hd\left(\hat{\modcw}, \modcw \right)
\end{align}
where $(a)$ holds if and only if $0<\epsilon<(M-1)/M$. Observe that the metric in \eqref{eq:MM_symbol} is equivalent to the optimum (i.e., \ac{ML}) metric for an $M$-ary symmetric channel with error probability $\epsilon$. Hence, a Hamming metric decoder is in fact treating the channel $\phxx$ as a symmetric channel, ignoring the information provided by the actual channel transition probabilities. An achievable rate (not necessarily the maximum one, though) under the mismatched decoding metric $\mm$ is given by the generalized mutual information (GMI) \cite{MM:Merhav94}
\begin{equation}
I_{\HDDSW}^{\mathsf{gmi}}=\sup_{s>0} \expect\left[ \log_2 \left( \frac{\mm(\hat{X},X)^s}{\frac{1}{M}\sum_{x\in\inalpha}\mm(\hat{X},x)^s}\right)  \right] \label{eq:Igmi_HDD}
\end{equation}
where the expectation is over $X,\hat{X}$.
We have the following result.

\begin{prop} The GMI for the symbol-wise standard hard decision decoding is 
	\begin{equation}
	I_{\HDDSW}^{\mathsf{gmi}}=\log_2 M - \hb(\es) - \es \log_2(M-1) \label{eq:Igmi_HDD_simp}
	\end{equation}
	where $\hb(\es)=-\es \log_2 \es - (1-\es)\log_2 (1-\es)$ is the binary entropy function evaluated in $\es$, $\es$ being the (uncoded) symbol error probability,
	$\es=\Pr\{\hat{X}\neq X\}$.
\end{prop}

\begin{IEEEproof} 
	We define 
	\begin{align}
	&I_{\HDDSW}^{\mathsf{gmi}}(s)\\
	&=\expect\left[ \log_2 \left( \frac{\mm(\hat{X},X)^s}{\frac{1}{M}\sum_{x\in\inalpha}\mm(\hat{X},x)^s}\right)  \right]\\
	&=\frac{1}{M}\sum_{\hat{x}\in\inalpha}\sum_{x\in\inalpha} \phxx(\hat{x}|x)\log_2 \left( \frac{M\mm(\hat{x},x)^s}{\sum_{x'\in\inalpha}\mm(\hat{x},x')^s}\right),
	\end{align}
	where in the second equality we used the fact that all symbols are equiprobable, i.e., $P_X(x)=1/M~$ for all $x$.
	
	By observing that 
	\begin{equation}
	\frac{\mm(\hat{x},x)^s}{\sum_{x'\in\inalpha}\mm(\hat{x},x')^s}= \frac{(1-\epsilon)^s}{(1-\epsilon)^s + (M-1)\left(\frac{\epsilon}{M-1}\right)^s}
	\end{equation}
	if $\hat{x}=x$ and
	\begin{equation}
	\frac{\mm(\hat{x},x)^s}{\sum_{x'\in\inalpha}\mm(\hat{x},x')^s}= \frac{\left(\frac{\epsilon}{M-1}\right)^s}{(1-\epsilon)^s + (M-1)\left(\frac{\epsilon}{M-1}\right)^s}
	\end{equation}
	if $\hat{x}\neq x$, we have that
	\begin{align}
	I_{\HDDSW}^{\mathsf{gmi}}(s)
	&=\log_2 M - (1-\es) \log_2 \left(1+ \frac{\epsilon^s (M-1)^{1-s}}{(1-\epsilon)^s}\right)\\
	&\quad-\es \log_2 \left(M-1+ \frac{(1-\epsilon)^s}{\epsilon^s (M-1)^{-s}}\right)
	\end{align}
	where we made use of the definition of the symbol error probability $\es=\Pr\{\hat{X}\neq X\}$. Since the choice of $\epsilon$ is arbitrary in $(0, (M-1)/M)$, we are allowed to pick $\epsilon=\es$.\footnote{Note that $\es$ is always upper bounded by $(M-1)/M$, which corresponds to the symbol error probability in the absence of channel observation.} By re-writing $I_{\HDDSW}^{\mathsf{gmi}}(s)$ as
	\begin{align}
	I_{\HDDSW}^{\mathsf{gmi}}&(s)=\log_2 M - (1-\delta) \log_2 \left(1+ (M-1)\Lambda^s\right)\\
	&-\es \log_2 \left(1+(M-1)^{-1}\Lambda^{-s}\right)-\es\log_2(M-1)
	\end{align}
	with $\Lambda\triangleq[\delta/(M-1)]/(1-\delta)$
	and setting its derivative in $s$ to $0$, we find that the optimum maximum of $I_{\HDDSW}^{\mathsf{gmi}}(s)$ is attained for $s=1$,  returning  \eqref{eq:Igmi_HDD_simp}.
\end{IEEEproof} 

\section{Achievable Information Rates with Bit-Wise Decoding}

Denote by $m=\log_2 M$ the number of bits used to label each constellation symbol, and by $\bm{b}=(b_1,b_2, \ldots, b_m)$ the $m$-bit labeling. 
We consider next binary coding schemes with different decoding strategies. For this purpose, we model the $m$ bit level channels as $m$ parallel independent \acp{BSC}, with the $i$th channel having bit error probability $\eb_i=\Pr\{\hat{B}_i\neq B_i\}$.
We consider the case where encoding is done across bit levels, i.e., BICM. With bit-wise HdChaD (HdChaD-BW) the decoder exploits the information on the different error probabilities associated with the $m$ bit levels. The GMI with HdChaD-BW is
\begin{align}\label{IBHSD}
I_{\mathsf{HdChaD-BW}}^{\mathsf{gmi}}=m - \sum_{i=1}^{m} \hb(\eb_i).
\end{align}
Note that the decoder exploits soft information, since the different reliabilities of the bit-wise hard decisions are embedded into the metric
\begin{equation}\label{hard_detec_bits}
P_{\hat{\bm{B}}|\bm{B}}(\hat{\bm{b}}|\bm{b})=\prod_{i=1}^m P_{\hat{B}_i|B_i}(\hat{b}_i|b_i)
\end{equation}
with
\begin{equation}
P_{\hat{B}_i|B_i}(\hat{b}_i|b_i)=\left\{ 
\begin{array}{ll}
1-\eb_i & \text{if } \hat{b}_i=b_i\\
\eb_i & \text{otherwise}.
\end{array}
\right.
\end{equation}
This rate corresponds to the bit-wise decoding in \cite{Lig16}\footnote{Note that in \cite{Lig16} it is implicitly assumed that the channel transitions probabilities of the resulting DMC are available at the decoder. Thus, in this paper we refer to this decoder as hard detection channel-aware decoder since the decoder indeed exploits the channel transition probabilities (i.e., soft information) in the decoding of the hard detected input.}.

Interestingly, this rate is also achievable by encoding bit levels separately (i.e., using multi-level coding), and decoding with the Hamming metric (without multi-stage decoding) \cite{MM:Lapidoth96}. 

On the other hand, by random coding across bit levels and bit-wise HDD (HDD-BW), i.e., Hamming metric decoding, is used at the receiver, the GMI is  
\begin{align}
I_{\HDDBW}^{\mathsf{gmi}}=m [1- \hb(\bar{\eb})]
\end{align}
where
\[
\bar{\eb}=\frac{1}{m}\sum_{i=1}^{m} \eb_i.
\]

\section{Numerical Results}\label{sec:studycase}

\subsection{Achievable Information Rates for the AWGN channel}

\begin{figure}[!t] 
\centering 
\includegraphics[width=\columnwidth]{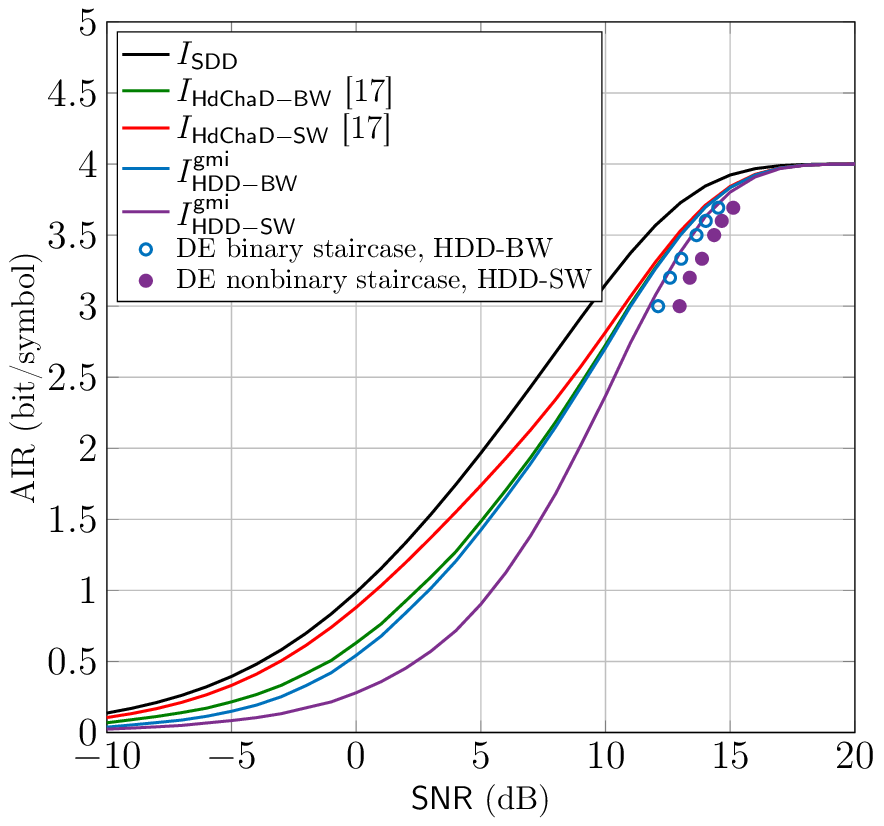}
	\vspace{-2ex}
	\caption{AIRs and DE performance of SC-GLDPC codes with HDD for $16$-QAM modulation. AWGN channel.} \vspace{-3ex}
	\label{16_QAM} 
\end{figure}
\begin{figure}[!t] \centering 
\includegraphics[width=\columnwidth]{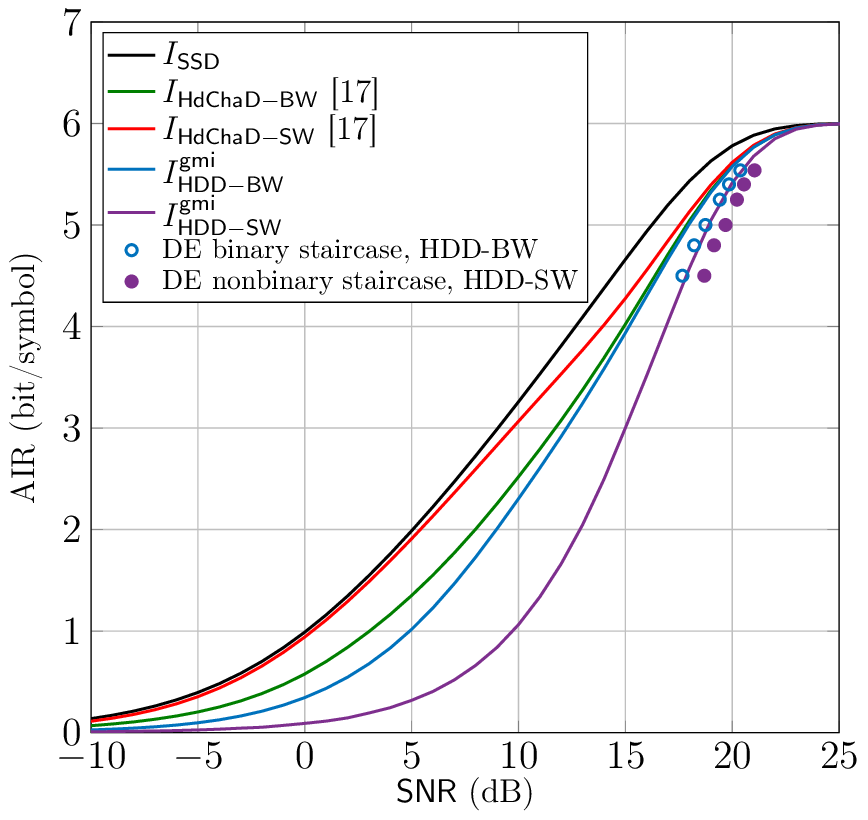}
	\vspace{-4ex}
	\caption{AIRs and DE performance of SC-GLDPC codes with HDD for $64$-QAM modulation. AWGN channel.} \vspace{1ex}
	\label{64_QAM} 
	\vspace{-5ex}
\end{figure}
\begin{figure}[!t] \centering 
\includegraphics[width=\columnwidth]{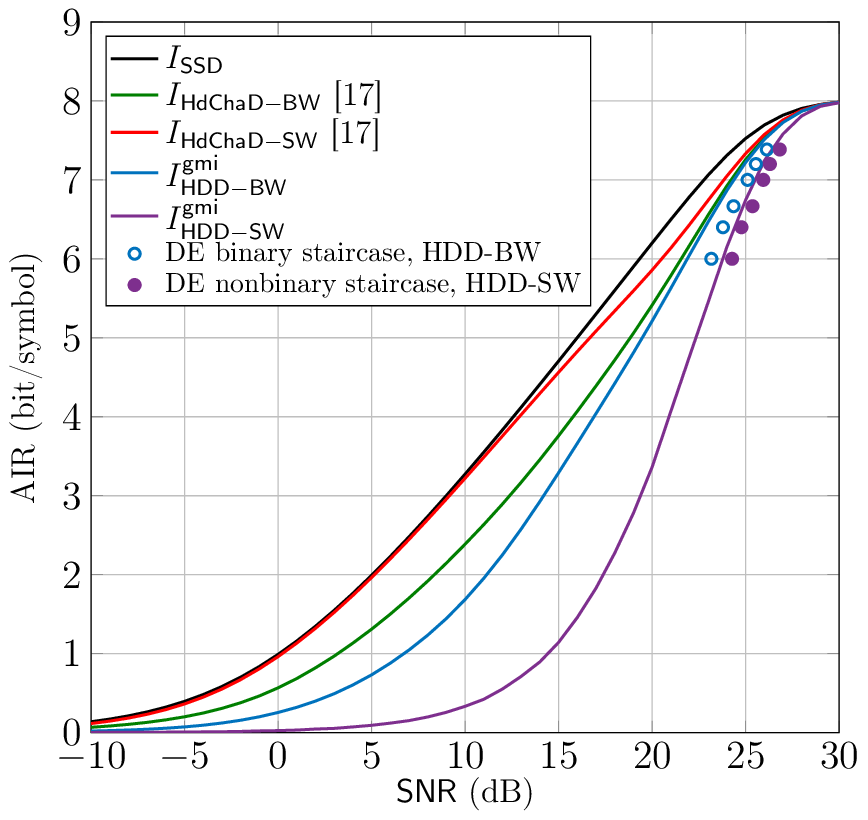}	
	\vspace{-5ex}
	\caption{AIRs and DE performance of SC-GLDPC codes with HDD for $256$-QAM modulation. AWGN channel.} \vspace{-1ex}
	\label{256_QAM} 
\end{figure}

In this section, we compare the AIRs for the different detection/decoding strategies discussed in the previous sections for the AWGN channel. In Fig.~\ref{16_QAM}, Fig.~\ref{64_QAM}, and Fig.~\ref{256_QAM} we plot the AIRs for $16$-QAM, $64$-QAM, and $256$-QAM, respectively, as a function of the SNR. It can be observed that, as already shown in \cite{Lig16}, with HdChaD the AIRs for the symbol-wise decoder are higher than those for the bit-wise decoder and close to those for SDD for low-to-medium SNRs. However, the behavior is completely the opposite if the conventional HDD is used. In this case, the AIRs for the bit-wise decoder are higher than those for the symbol-wise decoder. The penalty incurred by using the symbol-wise decoder is significant for low-to-medium SNRs. Equivalently, to achieve a given spectral efficiency, the symbol-wise decoder requires a significantly higher SNR. For example, for $64$-QAM and spectral efficiency $3$ bit/symbol the required SNR is $2$ dB higher for the symbol-wise decoder as compared to the bit-wise decoder. In the region of interest, corresponding to spectral efficiencies close to the saturation value ($6$ bit/symbol for $64$-QAM), the AIRs for the bit-wise and symbol-wise decoders get closer to each other, but the gain provided by the bit-wise decoder is non negligible. These results indicate that, for HDD, which is the decoding algorithm used in practice due to its low-complexity efficient implementation, binary FEC codes are preferable over nonbinary FEC codes.


To confirm these findings, we consider several binary and nonbinary codes and plot their performance in Figs.~\ref{16_QAM}--\ref{256_QAM}. For bit-wise HDD we consider binary staircase codes with BCH codes as component codes \cite{Smith12}. In particular, we consider the staircase codes designed in \cite{Hager15b}, which are optimized for each code overhead (OH) based on DE. DE is a tool to predict the iterative decoding performance of asymptotically long codes in the waterfall region. In particular, staircase codes can be seen as a particular class of spatially-coupled generalized LDPC codes (SC-GLDPCs) \cite{Hager15b}, thus it is possible to use DE, a well-established tool to analyze the performance of LDPC codes, to optimize the so-called decoding threshold, defined as the pre-FEC BER value where the DE curve crosses a certain target post-FEC BER when the code length grows to infinity. For HDD-SW we consider nonbinary staircase codes with RS codes as component codes. Nonbinary staircase codes were considered by the authors in \cite{She17}, where codes for different OHs were optimized based on DE. In Appendix~\ref{APPDE}, we derive the DE for nonbinary staircase codes (not published in \cite{She17} due to space limitations). In the figures, we plot the asymptotic performance of binary and nonbinary staircase codes, in the limit of very large block lengths, computed using DE. The DE assumes iterative bounded distance decoding with extrinsic message passing (EMP) for both binary and nonbinary staircase codes \cite{PfisterGC13,Henry_approachingcapacity_12,Hager15b}. Iterative BDD is the (suboptimal) low-complexity decoding algorithm used to decode staircase codes. Despite being suboptimal, it yields excellent performance. The empty circles correspond to the BER performance of binary staircase codes and the filled circles to the performance of nonbinary staircase codes. In all figures, for both binary and nonbinary staircase codes, starting from the circles at the top, the circles correspond to codes with OHs $8.33\%,11.11\%,14.29\%,20\%,25\%$, and $33.33\%$, respectively. It can be observed that the asymptotic performance of the staircase codes follow nicely the behavior predicted by the corresponding AIR curves. Also, it can be observed that the gap to the AIRs decreases for smaller OHs. This is expected, since staircase codes are known to perform particularly close to the theoretical limit for very high rates (corresponding to small OHs). Although perhaps difficult to observe in the figures, the binary staircase codes achieve gains between $0.6$ dB and $1.11$ dB with respect to the nonbinary counterparts. These results confirm the conclusions arising from the AIRs analysis, i.e., for spectrally-efficient systems, if HDD is used, binary codes should be preferred to nonbinary codes.
\begin{figure}[!t] \centering 
\includegraphics[width=\columnwidth]{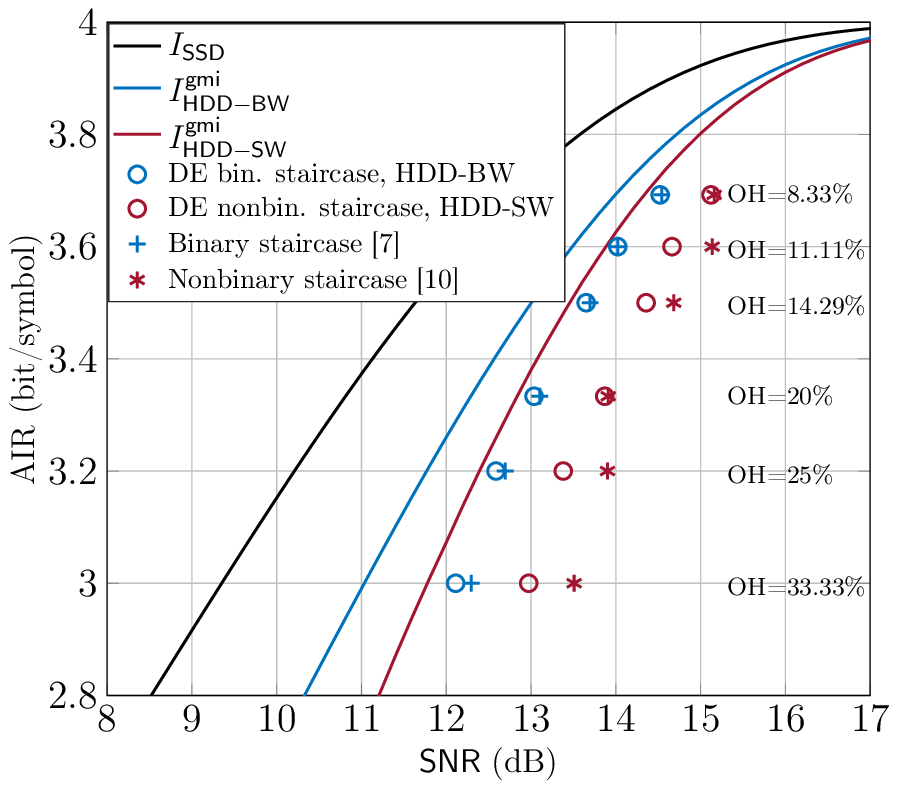}	 
	\vspace{-4ex}
	\caption{AIRs, DE performance of SC-GLDPC codes, and finite length performance of staircase codes with HDD for $16$-QAM modulation. AWGN channel.} 
	\vspace{-1ex}
	\label{16_QAM_real_codes} 
\end{figure}

In Fig.~\ref{16_QAM_real_codes}, we also plot the actual performance of several staircase codes (crosses for binary staircase codes and asterisks for nonbinary staircase codes). Both binary and nonbinary staircase codes are decoded using iterative BDD. For binary staircase codes, EMP  \cite{Henry_approachingcapacity_12} is used, which yields some performance gains with respect to the intrinsic message passing (IMP) algorithm proposed originally in \cite{Smith12} at the expense of a slight increase in complexity. For nonbinary staircase codes,  IMP is used, since EMP is highly complex.  Also, for both binary and nonbinary staircase codes, we used window decoding with a window size of $7$ staircase blocks and $4$ decoding iterations within the window. Comparing the finite length performance of binary and nonbinary staircase codes with the DE results, we observe that binary staircase codes perform very close to the predicted DE values. However, for some OHs, the use of IMP for nonbinary staircase codes entails a loss with respect to the performance predicted by DE, which results in an increased gap with respect to the performance of the binary staircase codes. Overall, the simulation results for finite code length also confirm the advantage of using binary codes.
\begin{figure*}%
	\centering
	\begin{subfigure}{6cm}
		\includegraphics{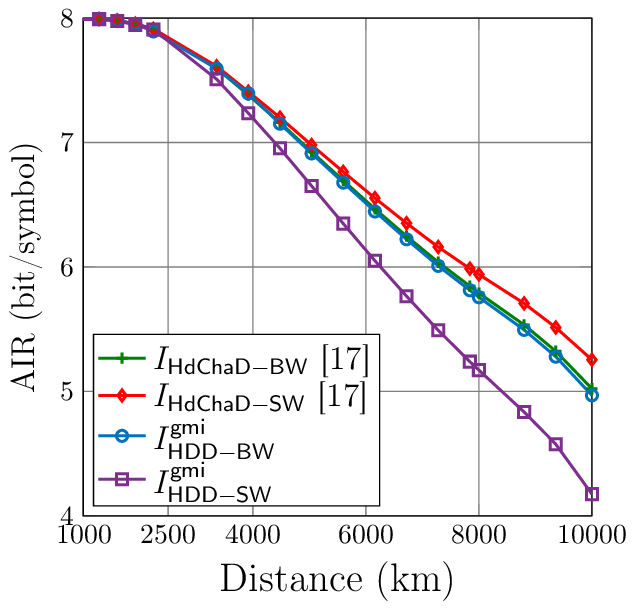}	  
		\vspace{-3ex}
		\caption{16-QAM} \vspace{1ex}
		\label{16_QAM_EDC_singlechannel} 
	\end{subfigure}\hfill%
	\begin{subfigure}{6cm}
         \includegraphics{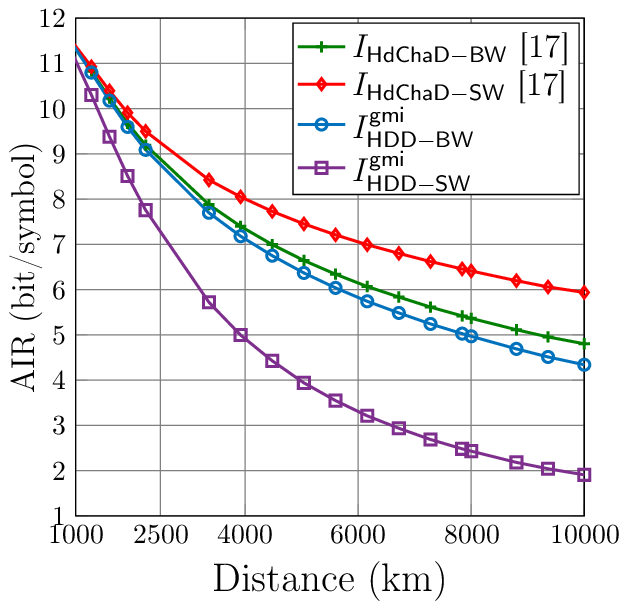}		
		\vspace{-3ex}
		\caption{64-QAM} \vspace{1ex}
		\label{64_QAM_EDC_singlechannel} 
	\end{subfigure}\hfill%
	\begin{subfigure}{6cm}
         \includegraphics{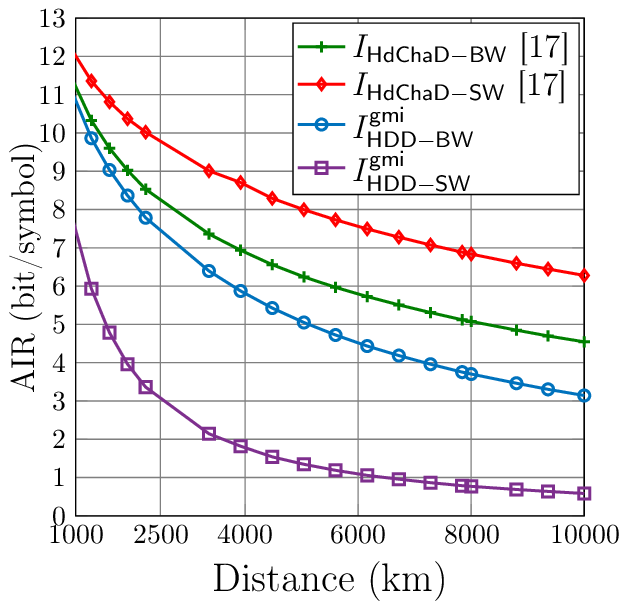}			 
		\vspace{-3ex}
		\caption{256-QAM} \vspace{1ex}
		\label{256_QAM_EDC_singlechannel} 
	\end{subfigure}%
	\caption{AIRs for the PM single channel transmission system with EDC as a function of the transmission distance.}
	\label{figEDC}
	\vspace{-1.2cm}
\end{figure*}
\hfill
\begin{figure*}%
	\centering
	\begin{subfigure}{6cm}
         \includegraphics{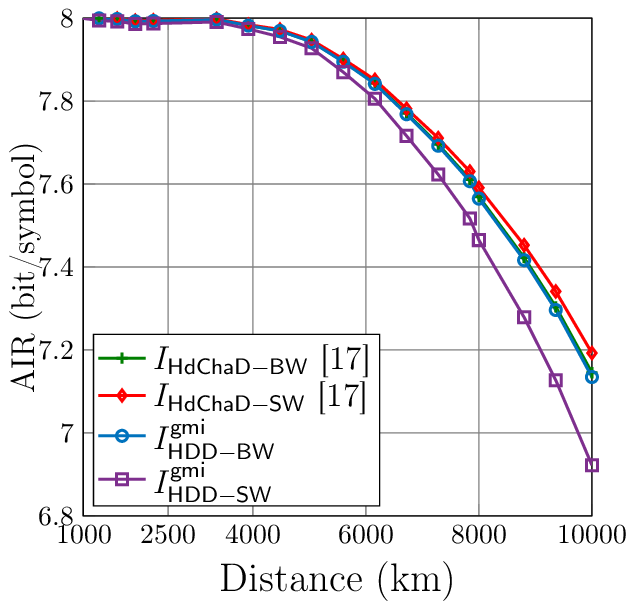}		
		\vspace{-3ex}
		\caption{16-QAM} \vspace{1ex}
		\label{16_QAM_DBP_singlechannel} 
	\end{subfigure}\hfill%
	\begin{subfigure}{6cm}
         \includegraphics{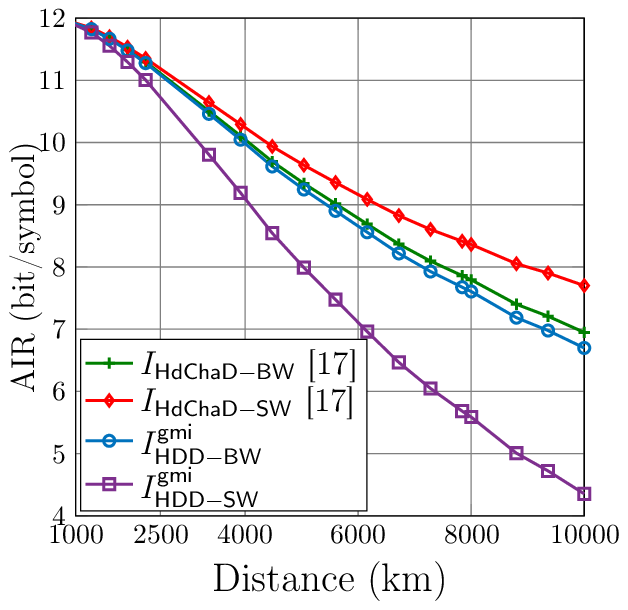}		
		\vspace{-3ex}
		\caption{64-QAM} \vspace{1ex}
		\label{64_QAM_DBP_singlechannel} 
	\end{subfigure}\hfill%
	\begin{subfigure}{6cm}
         \includegraphics{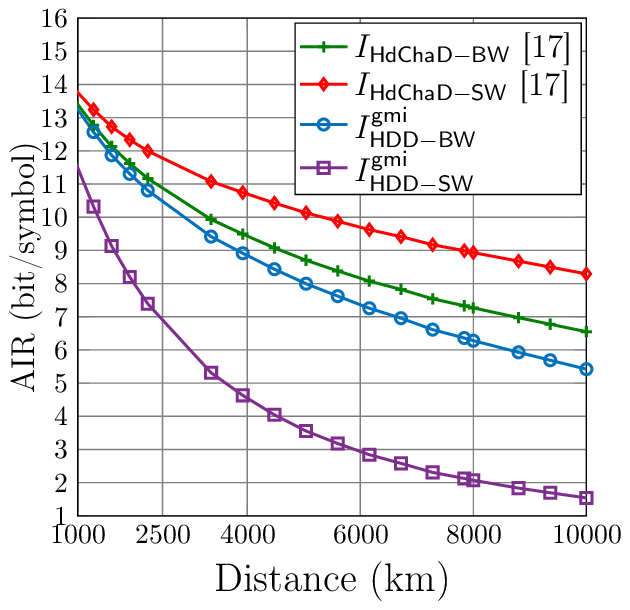}		 
		\vspace{-3ex}
		\caption{256-QAM} \vspace{1ex}
		\label{256_QAM_DBP_singlechannel} 
	\end{subfigure}%
	\caption{AIRs for the PM single channel transmission system with DBP as a function of the transmission distance.}
	\label{figDBP}
	\vspace{-1.2cm}
\end{figure*}
\hfill
\begin{figure*}%
	\centering
	\begin{subfigure}{6cm}
         \includegraphics{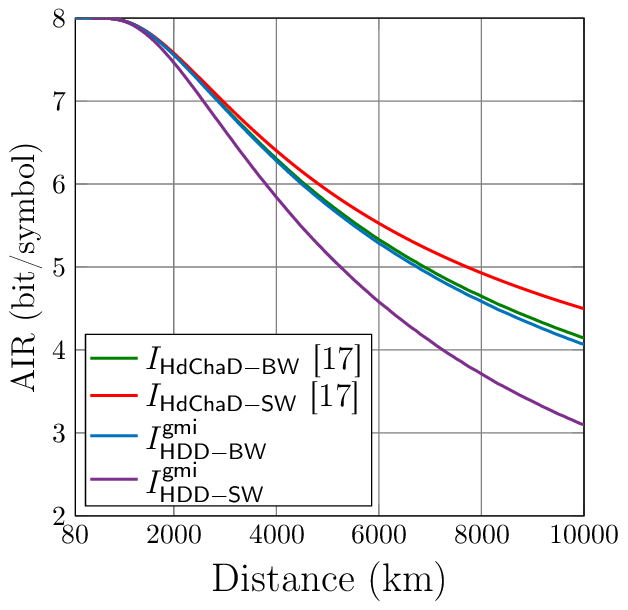}		 
		\vspace{-3ex}
		\caption{16-QAM} \vspace{1ex}
		\label{16_QAM_GN_singlechannel} 
	\end{subfigure}\hfill%
	\begin{subfigure}{6cm}
         \includegraphics{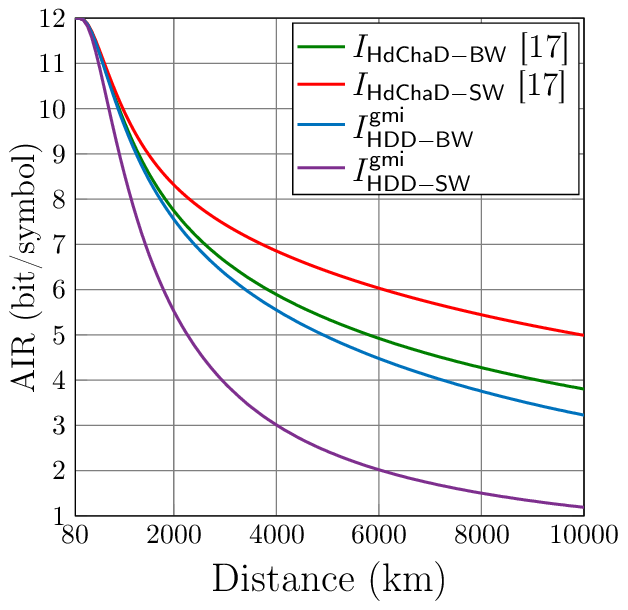}			
		\vspace{-3ex}
		\caption{64-QAM} \vspace{1ex}
		\label{64_QAM_GN_singlechannel} 
	\end{subfigure}\hfill%
	\begin{subfigure}{6cm}
         \includegraphics{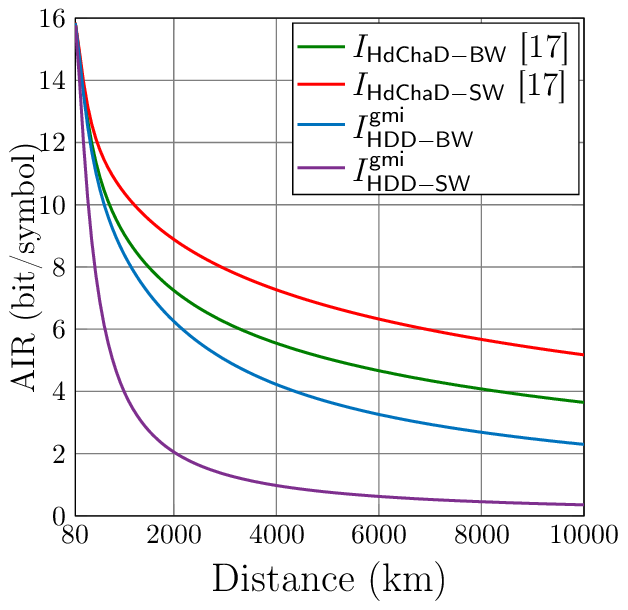}				
		\vspace{-3ex}
		\caption{256-QAM} \vspace{1ex}
		\label{256_QAM_GN_singlechannel} 
	\end{subfigure}%
	\caption{AIRs for the WDM system using the GN model as a function of the transmission distance.}
	\label{figGN}
\end{figure*}

%To verify different AIRs in a the optical channel, we have simulated the polarization multiplexed single channel transmission in the optical channel using split-step Fourier method to solve the Manakov equation. In particular, we considered a fiber with loss parameter $\alpha=0.2$ dB/km, dispersion parameter $D=17 $ ps/nm/km, nonlinear coefficient $\gamma=1.3$ $(\text{w.km})^{-1}$ and center wavelength $\lambda=1550$ nm. A single channel with $32$ Gbaud rate for each polarization is used for the simulation in a fiber with $80$ km spanlength where it is assumed that erbium doped fiber amplifier with noise figure $4.5$ dB compensates the loss perfectly in each span.

\subsection{Achievable Information Rates for the Fiber-Optic Channel}

In this section, we compute the AIRs for the fiber-optic channel, to verify that the conclusions drawn for the AWGN channel, i.e., bit-wise decoders perform better than symbol-wise decoders, hold also for the fiber-optic channel. In particular, we consider a PM single channel transmission system and a WDM transmission system. For the former, we assume an optical fiber with parameters summarized in Table~\ref{T1}. The span loss is compensated using erbium-doped fiber amplifiers (EDFAs). For the simulation of the fiber-optic channel, we used the SSFM to solve the Manakov equation \cite[eq. 3]{DBP}. To compensate the chromatic dispersion, we consider both electronic dispersion compensation (EDC) and digital back propagation (DBP). A root-raised cosine pulse with a roll-off factor of $0.25$ is used and for each polarization independent sequences of $2^{18}$ symbols are sent to compute the transition probabilities $\phxx(\hat{X}|X)$, ${\eb_i}$, $i=1,\cdots,m$, $\bar{\eb}$, and $\delta$, and correspondingly calculate the AIRs $I_{\mathsf{HdChaD-SW}}$, $I_{\mathsf{HdChaD-BW}}$,  $I_{\HDDBW}^{\mathsf{gmi}}$, and $I_{\HDDSW}^{\mathsf{gmi}}$. We also remark that for each span length, we used the optimal transmitted power found based on simulations.
\begin {table}[!t]
\renewcommand{\tabcolsep}{0.15cm}
\caption {Fiber and simulation parameters for the SSFM} \label{T1}
%\vspace{-0.25cm}
\begin{center}
	%\scriptsize
	\vspace{-2ex}
	\scriptsize
	\begin{center}\begin{tabular}{cccc}
			\arrayrulecolor{black}\hline
			\toprule

			Attenuation ($\alpha$) & Dispersion ($D$) & Nonlinear coefficient ($\gamma$) & $\lambda$ \\
%			(dB/km) & (ps/nm/km) & 1/(W.km)  & (nm) \\			
			\midrule
            $0.2$ dB/km & $17$ ps/nm/km & $1.3$ 1/(W.km) & $1550$ nm\\
			\toprule
			Symbol rate & Spanlength & EDFA noise figure & SSFM step size \\
%			(Gbaud) & (km) & (dB)  & (km) \\	
			\midrule
			$32$ Gbaud & $80$ km & $4.5$ dB & $0.1$ km\\						
			\toprule                			
		\end{tabular} \end{center}
	\end{center}
	\vspace{-0.5cm}
\end{table}

In Fig.~\ref{figEDC} and Fig.~\ref{figDBP}, we plot the AIRs for the PM single channel transmission system with EDC and DBP, respectively, for different modulations as a function of the transmission distance. As can be seen, in all cases, for HDD bit-wise decoding achieves larger AIRs compared to symbol-wise decoding, which in turn leads to a significant optical reach enhancement. %Furthermore, as expected, performing the DBP improves different AIRs leading also to the optical reach enhancement compared to the EDC, at the cost of increasing the receiver complexity.

For the sake of completeness, we also consider a WDM transmission system with $81$ channels and the same parameters summarized in Table~\ref{T1}. The optical channel for such a system is well-approximated by the AWGN channel, a model known as the GN model in the literature\cite{johannisson_2014_jlt}. All channels are transmitted with the same power, where the optimal power is found to maximize the SNR of the middle channel. The AIRs for the middle channel with $16$-QAM, $64$-QAM, and $256$-QAM are shown in Fig.~\ref{figGN}. The AIR of the middle channel can be interpreted as the pessimistic approximation of the AIRs of the other channels, since the middle channel suffers more from crosstalk. As for the AWGN channel and the PM single channel system, if HDD is considered, bit-wise decoding achieves a significantly larger optical reach in comparison to symbol-wise decoding, leading to the conclusion that binary/codes/bit-wise decoders are preferable than nonbinary codes/symbol-wise decoders. In particular, the optical reach enhancement of bit-wise decoding compared to symbol-wise decoding for $16$-QAM and $6$ bits/symbol, $64$-QAM and $8$ bits/symbol, and $256$-QAM and $10$ bits/symbol, is $701$ km, $590$ km, and $311$ km, respectively.

%In particular, we considered a fiber with loss parameter $\alpha=0.2$ dB/km, dispersion parameter $D=17 $ ps/nm/km, nonlinear coefficient $\gamma=1.3$ $(\text{w.km})^{-1}$ and center wavelength $\lambda=1550$ nm. A single channel with $32$ Gbaud rate for each polarization is used for the simulation in a fiber with $80$ km spanlength where it is assumed that erbium doped fiber amplifier with noise figure $4.5$ dB compensates the loss perfectly in each span.

\section{Conclusion}

We analyzed the achievable information rates of coded modulation schemes decoded using the standard hard decision decoder, which is based on the Hamming distance metric, for both bit-wise and symbol-wise decoders. We showed that for HDD, the AIRs of bit-wise decoders are higher than those of symbol-wise decoders. We also designed binary and nonbinary staircase codes and showed that their performance are in agreement with the behavior predicted by the AIRs. An interesting outcome of this work is that, if HDD is used, binary codes are preferable to nonbinary codes, as they achieve significantly better performance. Furthermore, since the decoding complexity of nonbinary codes is higher than that of binary codes, there seems to be no reason for considering nonbinary codes for HDD.

\section*{Acknowledgments}
The authors would like to thank Fabian Steiner, Georg B\"{o}cherer and Patrick Schulte (TUM-LNT) for their helpful comments and for identifying an inconsistency with the constellation labeling in an early version of the manuscript. 

\appendices
\section{Density Evolution for Nonbinary GLDPC/SC-GLDPC Codes}\label{APPDE}

In this appendix, we derive the DE for nonbinary staircase codes with RS codes of length $n$ as component codes. We assume that the codes are decoded using HDD, i.e., the received symbols are mapped to the nearest constellation point, and decoding is performed iteratively assuming BDD of the component codes. As explained in \cite{She17}, nonbinary staircase codes can be seen as a subclass of the ensemble of nonbinary SC-GLDPC codes. A nonbinary SC-GLDPC code can be represented by a bipartite graph consisting of variable nodes (VNs), corresponding to code bits, and constraint nodes (CNs). The SC-GLDPC code ensemble is defined by the parameters $(\mathcal{C},L,w)$, where $L$ is the number of spatial positions, and $w$ is the coupling width. Nonbinary staircase codes are then contained in the SC-GLDPC code ensemble when the CNs correspond to RS codes and $w=2$, i.e., each coded symbol is protected by two component codes \cite{She17}. This allows us to optimize the parameters of nonbinary staircase codes using DE. DE tracks the evolution of the average symbol error probability in the iterative BDD algorithm.

\begin{figure}[t!]
	\centering
	 \includegraphics[width=0.7\columnwidth]{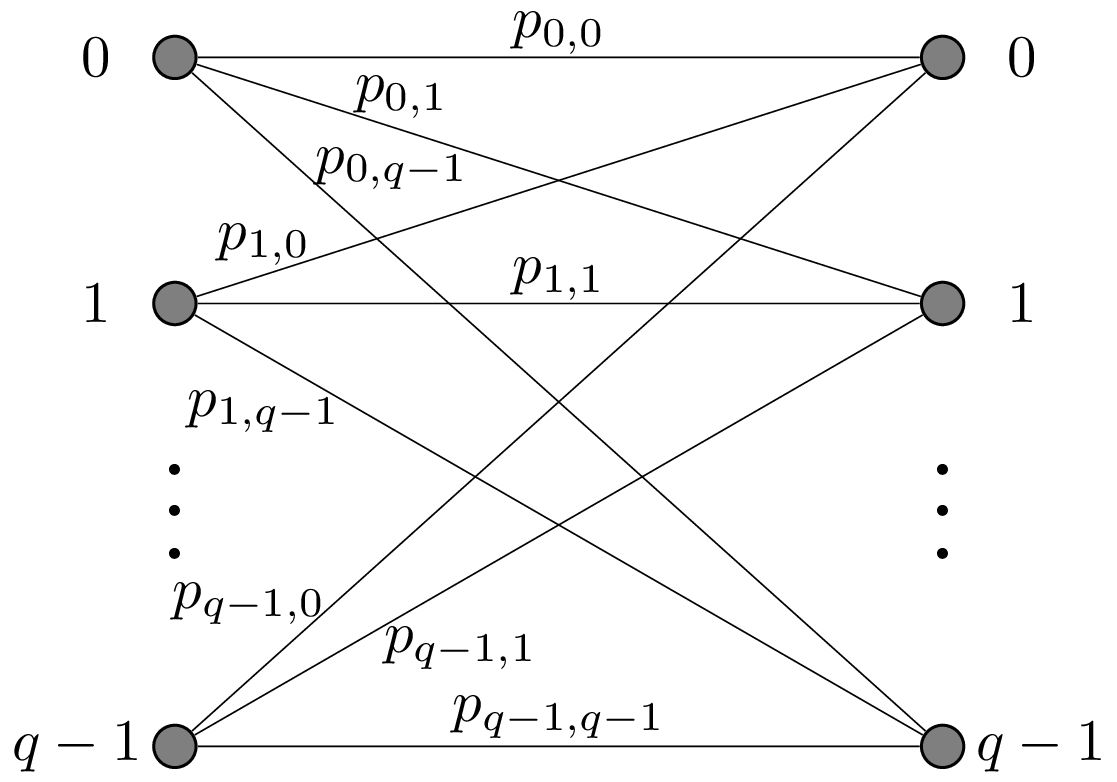}	
	\caption{The $q$-ary input $q$-ary output channel.}
	\vspace{-0.4cm}
	\label{fig:regulargraph}
\end{figure}

Assume that the nonbinary staircase code is constructed over a Galois field matched to the constellation size, i.e., the coded symbols from GF$(q)$ are mapped to a $q$-QAM constellation. As already mentioned in Section~\ref{sec:prelim}, the optical channel can be modeled as an AWGN channel using the GN model under certain conditions \cite{Pog12}. HDD transforms the AWGN channel into a $q$-ary input $q$-ary output channel, referred to as the $q$-ary channel in the following and depicted in Fig.~\ref{fig:regulargraph}. The simplest $q$-ary channel is the $q$-ary symmetric channel (QSC), where symbol $i$ is received correctly with probability $p_{i,i}=1-p$ for $i=0,1,...,q-1$, and it is mistaken onto any other symbol with the same probability, i.e., $p_{i,j}=\frac{p}{q-1}$ for $i \ne j, \;i,j=0,1,...,q-1$. Here, we model the nonbinary transmission channel as a QSC. We remark that the AWGN channel deviates from the QSC as $q$ increases. However, one can interpret the QSC as an auxiliary channel for the true AWGN channel, which simplifies the code design, since the symmetry of the channel and the decoder allows to make the simplifying assumption that the all-zero codeword is transmitted, i.e., the analysis is codeword-independent. More importantly, the use of the QSC is justified by the fact that, as shown in Section~\ref{sec:AIRsHDDsw}, the HDD treats indeed the channel as a symmetric channel. As a result, the best nonbinary codes arising from the DE analysis derived below for the QSC perform also the best for the AWGN channel. 

In the following, we generalize the DE for binary SC-GLDPC codes over the binary symmetric channel in \cite{Henry_approachingcapacity_12} to the nonbinary SC-GLDPC code ensemble over the QSC, rigorously accounting for decoding miscorrections. For ease of exposition, we consider first GLDPC codes and then SC-GLDPC codes. 
Let $x^{(\ell)}$ be the symbol error probability from VNs to CNs in the $\ell$-th iteration, where $x^{(0)}=p$. Furthermore, let ${P_n}\left( i \right)$ be the probability that a randomly selected symbol (out of the $n$ code symbols of the component code) is decoded erroneously when prior to decoding it was also in error, and there are $i$ other symbol errors in the remaining $n-1$ code symbols of the component code. Similarly, denote by ${{\bar P}_n}\left( i \right)$ the probability that a randomly selected symbol (out of the $n$ code symbols of the component code) is decoded erroneously when prior to decoding it was not in error and there are $i$ other symbols in error among the remaining $n-1$ code symbols of the component code. With this notation, ${{\bar P}_n}\left( i \right)$ accounts for the miscorrections. 

It can be shown that $x^{(\ell+1)}=f(x^{(\ell)};p_s)$, where
\begin{align}\label{fx} 
\nonumber {f}(x;{p_s}) \buildrel \Delta \over = \sum\limits_{i = 0}^{n - 1} &  {\Big( {\begin{array}{*{20}{c}}
		{n - 1}\\
		i
		\end{array}} \Big)} {\Big( {\frac{x}{{q - 1}}} \Big)^i}{\left( {q - 1} \right)^i}{\left( {1 - x} \right)^{n - i - 1}} \cdot \\  & \Big( {\Big( {\sum\limits_{j = 1}^{q - 1} {\Big( {\frac{{{p_s}}}{{q - 1}}} \Big)} } \Big){P_n}\left( i \right) + {{\bar p}_s}{{\bar P}_n}\left( i \right)} \Big).
\end{align}
The function $f(x;p_s)$ can be simplified as
\begin{align}\label{fx_simple} 
{f}(x;{p_s}) =  \sum\limits_{i = 0}^{n - 1} & {\Big( {\begin{array}{*{20}{c}}
		{n - 1}\\
		i
		\end{array}} \Big)}{\left( {1 - x} \right)^{n - i - 1}} {x^i} \cdot \\ & \nonumber \Big( {{p_s}{P_n}\left( i \right) + {{\bar p}_s}{{\bar P}_n}\left( i \right)} \Big),
\end{align}
where
\begin{align}
\label{P}
{P_n}\left( i \right) = \begin{dcases}
1 - \mathop {\sum\limits_{\delta  = 1}^t {\sum\limits_{j = 0}^{\delta  - 1} {\sum\limits_{z = 0}^j {} } } } \mathds{1}_{\alpha} \frac{{n - \alpha}}{n}{A_\alpha}F_{\alpha} &
\text{if} \;\; t\le i\\
0 &
\text{if} \;\;  0 \le i \le t - 1  \\
\end{dcases}
\end{align} 
and
\begin{align}
\label{Pbar}
{{\bar P}_n}\left( i \right) = \begin{dcases}
\mathop {\sum\limits_{\delta  = 1}^t {\sum\limits_{j = 0}^{\delta  - 1} {\sum\limits_{z = 0}^j {} } } } \mathds{1}_{\alpha} \frac{{\alpha + 1}}{n}{A_{\alpha+ 1}}F_{\alpha} &
\text{if} \;\; t+1 \le i\\
0 &
\text{if} \;\;  0 \le i \le t  \\
\end{dcases}
\end{align}
with $\alpha = i - \delta  + 2j - z + 1$,
\begin{align}
\mathds{1}_{\alpha} =
\begin{dcases}
1 &
\text{if} \;\; {n - \alpha - 1 \ge \delta  - j - 1}\\
0 &
\text{otherwise}  \\
\end{dcases},
\end{align}
and
\begin{align}\label{f_w} 
\small{F_{\alpha} = \frac{{\Big( {\begin{array}{*{20}{c}}
				{\alpha}\\
				{\alpha - j}
				\end{array}} \Big)\Big( {\begin{array}{*{20}{c}}
				j\\
				z
				\end{array}} \Big){{\left( {q - 2} \right)}^z}\Big( {\begin{array}{*{20}{c}}
				{n - \alpha - 1}\\
				{\delta  - j - 1}
				\end{array}} \Big){{\left( {q - 1} \right)}^{\delta  - j - 1}}}}{{\Big( {\begin{array}{*{20}{c}}
				{n - 1}\\
				i
				\end{array}} \Big){{\left( {q - 1} \right)}^i}}}}.
\end{align} 

$A_\alpha$ in \eqref{P} and \eqref{Pbar} is the number of codewords of the component code of weight $w$. For RS codes constructed in $\text{GF}(q)$, $A_w$ is given as \cite[eq. 8]{Kasami_84}
\begin{align}\label{weight_dist} 
{A_\alpha} = & \left( {\begin{array}{*{20}{c}}
	{q - 1}\\
	\alpha
	\end{array}} \right){q^{ - 2t}}  {{\left( {q - 1} \right)}^i} +  \\ & \nonumber \left( {\begin{array}{*{20}{c}}
	{q - 1}\\
	\alpha
	\end{array}} \right){q^{ - 2t}}\sum\limits_{j = 0}^{2t} {{{\left( { - 1} \right)}^{i + j}}} \left( {\begin{array}{*{20}{c}}
	i\\
	j
	\end{array}} \right)\left( {{q^{2t}} - {q^j}} \right)\yesnumber .
\end{align} 
One can use \eqref{fx} to compute the decoding threshold of the GLDPC code with RS codes as component codes to achieve a target post-FEC BER.  

The derived DE equations can be easily extended to SC-GLDPC codes. For SC-GLDPC codes, we track the average symbol error probabilities exchanged in the iterative decoding for each spatial position. Each VN at position $i$ is connected to CNs at positions $[i-w+1,i]$. Let $x^{(\ell)}_i$ be the average symbol error probability from VNs at spatial position $i$ to the connected CNs at spatial positions in $[i-w+1,i]$. Also, let $x_{i,c}^{(\ell )}$ be the average symbol error probability from CNs at spatial position $i$ to VNs at positions $[i,i+w-1]$. $x^{(\ell)}_i$ and $x_{i,c}^{(\ell+1)}$ can be calculated as
\begin{align}\label{xl} 
x_{i,c}^{(\ell )} = \frac{1}{{{w}}}\sum\limits_{j = 0}^{{w} - 1} {x_{i + j}^{(\ell )}}, 
\end{align}
\begin{align}\label{xcl} 
x_i^{(\ell+1 )} = \frac{1}{{{w}}}\sum\limits_{l = 0}^{{w} - 1} {{f}\left( {x_{i - l,c}^{(\ell )},{p_s}} \right)}. 
\end{align}
Based on \eqref{xl} and \eqref{xcl}, $x^{(\ell+1)}_i$ can be computed recursively as 
\begin{align}\label{xl_recursive} 
x_i^{(\ell  + 1)} = \frac{1}{{{w}}}\sum\limits_{l = 0}^{{w} - 1} {{f}\left( {\frac{1}{{{w}}}\sum\limits_{j = 0}^{{w} - 1} {x_{i - l + j}^{(\ell )}} ,{p_s}} \right)}. 
\end{align}

In general, the decoder of SC-GLDPC codes (and therefore of staircase codes as well) is based on the sliding window decoding. One can easily modify the DE equations above to account for the window decoding. Assume that the width of the window is $W$, i.e., the window contains $W$ spatial positions. Let $\mathcal{W}_j$ denote the set containing the collection of VNs and CNs involved in the decoding of the \emph{j}th slided window. The decoder freezes the VNs and CNs outside the window, i.e., the VNs and CNs inside the window are updated based on the information exchanged inside the window and no information comes from the positions outside the window. Therefore, one can define $x^{\prime(\ell )}_i$ and $x^{\prime(\ell )}_{i,c}$ as
\begin{align}
\label{al_prim_window}
x^{\prime(\ell )}_i =
\begin{dcases}
0 &
\text{if } i \notin {{\cal W}_j}\\
x^{(\ell)}_i &
\text{if } i \in {{\cal W}_j} \\
\end{dcases},
\end{align} 
\begin{align}
\label{al_prim_c_window}
x_{i,c}^{\prime(\ell )} =
\begin{dcases}
0 &
\text{if } i \notin {{\cal W}_j}\\
x_{i,c}^{(\ell )} &
\text{if } i \in {{\cal W}_j} \\
\end{dcases},
\end{align} 
and use them in \eqref{xl} and \eqref{xcl} to find the average symbol error probability for the spatial positions within the window, i.e., 
\begin{align}\label{xl_recursive_window} 
x_i^{\prime(\ell  + 1)} = \frac{1}{{w}}\sum\limits_{l = 0}^{{w} - 1} {{f}\left( {\frac{1}{{w}}\sum\limits_{j = 0}^{{w} - 1} {x_{i - l + j}^{\prime(\ell )}} ,{p_s}} \right)}
\end{align}
for $i \in {{\cal W}_j}$. Finally, using \eqref{xl_recursive_window}, one can compute the average symbol error probability for each spatial position of the SC-GLDPC code with window decoding. The average symbol error probability over all spatial positions is then the post-FEC symbol error probability of the SC-GLDPC code.

\balance


\begin{thebibliography}{10}
	\providecommand{\url}[1]{#1}
	\csname url@samestyle\endcsname
	\providecommand{\newblock}{\relax}
	\providecommand{\bibinfo}[2]{#2}
	\providecommand{\BIBentrySTDinterwordspacing}{\spaceskip=0pt\relax}
	\providecommand{\BIBentryALTinterwordstretchfactor}{4}
	\providecommand{\BIBentryALTinterwordspacing}{\spaceskip=\fontdimen2\font plus
		\BIBentryALTinterwordstretchfactor\fontdimen3\font minus
		\fontdimen4\font\relax}
	\providecommand{\BIBforeignlanguage}[2]{{%
			\expandafter\ifx\csname l@#1\endcsname\relax
			\typeout{** WARNING: IEEEtran.bst: No hyphenation pattern has been}%
			\typeout{** loaded for the language `#1'. Using the pattern for}%
			\typeout{** the default language instead.}%
			\else
			\language=\csname l@#1\endcsname
			\fi
			#2}}
	\providecommand{\BIBdecl}{\relax}
	\BIBdecl
	
	\bibitem{Sheikh17}
	A.~Sheikh, A.~{Graell i Amat}, and G.~Liva, ``On achievable information rates
	for coherent fiber-optic systems with hard decision decoding,'' in
	\emph{Proc.\ Eur. Conf. Opt. Commun. (ECOC)}, Gothenburg, Sweden, Sep. 2017.
	
	\bibitem{Djordjevic2004}
	I.~B. Djordjevic and B.~Vasic, ``High code rate low-density parity-check codes
	for optical communication systems,'' \emph{IEEE Photon. Technol. Lett.},
	vol.~16, no.~6, pp. 1600--1602, Jun. 2004.
	
	\bibitem{Djordjevic_GLDPC}
	I.~B. Djordjevic, O.~Milenkovic, and B.~Vasic, ``Generalized low-density
	parity-check codes for optical communication systems,'' \emph{IEEE/OSA
		J.~Lightw. Technol.}, vol.~23, no.~5, pp. 1939--1946, May 2005.
	
	\bibitem{Hager15}
	C.~H\"{a}ger, A.~{Graell i Amat}, F.~Br\"{a}nnstr\"{o}m, A.~Alvarado, and
	E.~Agrell, ``Terminated and tailbiting spatially-coupled codes with optimized
	bit mappings for spectrally efficient fiber-optical systems,'' \emph{IEEE/OSA
		J.~Lightw. Technol.}, vol.~33, no.~7, pp. 1275--1285, Apr. 2015.
	
	\bibitem{Schmalen15}
	L.~Schmalen, V.~Aref, J.~Cho, D.~Suikat, D.~R\"osener, and A.~Leven,
	``Spatially coupled soft-decision forward error correction for future
	lightwave systems,'' \emph{IEEE/OSA J.~Lightw. Technol.}, vol.~33, no.~5, pp.
	1109--1116, Mar. 2015.
	
	\bibitem{Smith12}
	B.~P. Smith, A.~Farhood, A.~Hunt, F.~R. Kschischang, and J.~Lodge, ``Staircase
	codes: {FEC} for 100 {Gb}/s {OTN},'' \emph{J.~Lightw. Technol.}, vol.~30,
	no.~1, pp. 110--117, Jan. 2012.
	
	\bibitem{Hager15b}
	C.~H\"{a}ger, A.~{Graell i Amat}, H.~D. Pfister, A.~Alvarado,
	F.~Br\"{a}nnstr\"{o}m, and E.~Agrell, ``On parameter optimization for
	staircase codes,'' in \emph{Proc.\ {OSA} Optical Fiber Communication Conf.
		and Exhibition (OFC)}, Mar. 2015.
	
	\bibitem{PfisterGC13}
	Y.-Y. Jian, H.~D. Pfister, K.~R. Narayanan, R.~Rao, and R.~Mazahreh,
	``Iterative hard-decision decoding of braided {BCH} codes for high-speed
	optical communication,'' in \emph{Proc.~IEEE Global Telecom. Conf.
		(GLOBECOM)}, Dec. 2013.
	
	\bibitem{Hag16b}
	C.~H\"ager, A.~{Graell i Amat}, H.~D. Pfister, and F.~Br\"annstr\"om, ``Density
	evolution for deterministic generalized product codes with higher-order
	modulation,'' in \emph{Proc. 9th Int. Symp. Turbo Codes \& Iterative Inf.
		Processing (ISTC)}, Sep. 2016, pp. 236--240.
	
	\bibitem{She17}
	A.~Sheikh, A.~{Graell i Amat}, and M.~Karlsson, ``Nonbinary staircase codes for
	spectrally and energy efficient fiber-optic systems,'' in \emph{Proc.\ {OSA}
		Optical Fiber Communication Conf. and Exhibition (OFC)}, Mar. 2017.
	
	\bibitem{Djordjevic2005}
	I.~B. Djordjevic, B.~Vasic, M.~Ivkovic, and I.~Gabitov, ``Achievable
	information rates for high-speed long-haul optical transmission,''
	\emph{IEEE/OSA J.~Lightw. Technol.}, vol.~23, no.~11, pp. 3755--3763, Nov.
	2005.
	
	\bibitem{arnold2006}
	D.~M. Arnold, H.~A. Loeliger, P.~O. Vontobel, A.~Kavcic, and W.~Zeng,
	``Simulation-based computation of information rates for channels with
	memory,'' \emph{IEEE Trans. Inf. Theory}, vol.~52, no.~8, pp. 3498--3508,
	Aug. 2006.
	
	\bibitem{Ess10}
	R.~J. Essiambre, G.~Kramer, P.~J. Winzer, G.~J. Foschini, and B.~Goebel,
	``Capacity limits of optical fiber networks,'' vol.~28, no.~4, pp. 662--701,
	Feb. 2010.
	
	\bibitem{Kaplan1993}
	G.~Kaplan and S.~S. (Shitz), ``Information rates and error exponents of
	compound channels with application to antipodal signaling in a fading
	environment,'' \emph{AE{\"U} (Electron. and Commun.)}, vol.~47, no.~4, pp.
	228--239, Nov. 1993.
	
	\bibitem{Sec13}
	M.~Secondini, E.~Forestieri, and G.~Prati, ``Achievable information rate in
	nonlinear {WDM} fiber-optic systems with arbitrary modulation formats and
	dispersion maps,'' vol.~31, no.~23, pp. 3839--3852, Dec. 2013.
	
	\bibitem{Fehenberger_15}
	T.~Fehenberger, A.~Alvarado, P.~Bayvel, and N.~Hanik, ``On achievable rates for
	long-haul fiber-optic communications,'' \emph{Opt. Express}, vol.~23, no.~7,
	pp. 9183--9191, Apr. 2015.
	
	\bibitem{Lig16}
	G.~Liga, A.~Alvarado, E.~Agrell, and P.~Bayvel, ``Information rates of
	next-generation long-haul optical fiber systems using coded modulation,''
	\emph{J.\ Lightw.\ Technol.}, vol.~35, no.~1, pp. 113--123, Jan. 2017.
	
	\bibitem{LinCos04}
	S.~Lin and D.~J. {Costello Jr.}, \emph{Error Control Coding, Second
		Edition}.\hskip 1em plus 0.5em minus 0.4em\relax Upper Saddle River, NJ, USA:
	Prentice-Hall, Inc., 2004.
	
	\bibitem{lapidoth17}
	A.~Lapidoth, \emph{A foundation in digital communication}.\hskip 1em plus 0.5em
	minus 0.4em\relax Cambridge, UK: Cambridge University Press, 2009.
	
	\bibitem{barg1997complexity}
	A.~Barg, ``Complexity issues in coding theory,'' \emph{Handbook of Coding
		Theory}, vol. I, V.S. Pless, W.C. Huffman, Editors, North Holland, 2007.
	
	\bibitem{guruswami2004list}
	V.~Guruswami, \emph{List decoding of error-correcting codes}.\hskip 1em plus
	0.5em minus 0.4em\relax Springer Science \& Business Media, 2004.
	
	\bibitem{vardy1997algorithmic}
	A.~Vardy, ``Algorithmic complexity in coding theory and the minimum distance
	problem,'' in \emph{Proc. 29th Annual ACM Symp. Theory of Computing}, May
	1997, pp. 92--109.
	
	\bibitem{Pog12}
	P.~Poggiolini, ``The {GN} model of non-linear propagation in uncompensated
	coherent optical systems,'' \emph{J. Lightw. Technol.}, vol.~30, no.~24, pp.
	3857--3879, Dec. 2012.
	
	\bibitem{Schmalen2017}
	L.~Schmalen, A.~Alvarado, and R.~Rios-Müller, ``Performance prediction of
	nonbinary forward error correction in optical transmission experiments,''
	\emph{IEEE/OSA J.~Lightw. Technol.}, vol.~35, no.~4, pp. 1015--1027, Feb.
	2017.
	
	\bibitem{Henry_approachingcapacity_12}
	Y.~Y. Jian, H.~D. Pfister, and K.~R. Narayanan, ``Approaching capacity at high
	rates with iterative hard-decision decoding,'' in \emph{Proc. Int. Symp. Inf.
		Theory ({ISIT})}, Jul. 2012, pp. 2696--2700.
	
	\bibitem{MM:Merhav94}
	N.~Merhav, G.~Kaplan, A.~Lapidoth, and S.~{Shamai (Shitz)}, ``On information
	rates for mismatched decoders,'' \emph{IEEE Trans.\ Inf.\ Theory}, vol.~40,
	no.~6, pp. 1953--1967, Nov. 1994.
	
	\bibitem{MM:Lapidoth96}
	A.~Lapidoth, ``Mismatched decoding and the multiple-access channel,''
	\emph{IEEE Trans.\ Inf.\ Theory}, vol.~42, no.~5, pp. 1439--1452, Sep. 1996.
	
	\bibitem{DBP}
	A.~Napoli, Z.~Maalej, V.~A. J.~M. Sleiffer, M.~Kuschnerov, D.~Rafique,
	E.~Timmers, B.~Spinnler, T.~Rahman, L.~D. Coelho, and N.~Hanik, ``Reduced
	complexity digital back-propagation methods for optical communication
	systems,'' \emph{IEEE/OSA J.~Lightw. Technol.}, vol.~32, no.~7, pp.
	1351--1362, Apr. 2014.
	
	\bibitem{johannisson_2014_jlt}
	P.~Johannisson and E.~Agrell, ``Modeling of nonlinear signal distortion in
	fiber-optic networks,'' \emph{IEEE/OSA J.~Lightw. Technol.}, vol.~32, no.~23,
	pp. 3942--3950, Dec. 2014.
	
	\bibitem{Kasami_84}
	T.~Kasami and S.~Lin, ``On the probability of undetected error for the maximum
	distance separable codes,'' \emph{IEEE Trans.\ Commun.}, vol.~32, no.~9, pp.
	998--1006, Sep. 1984.
	
\end{thebibliography}
\end{document}